\documentclass[a4paper,aps,pra,reprint,superscriptaddress, twocolumn,preprintnumbers,amsmath,amssymb,nobalancelastpage]{revtex4-1}
\pdfoutput=1
\usepackage[dvips]{graphics}
\usepackage{bm}
\usepackage{float}
\usepackage{epsfig}
\usepackage{enumerate}
\usepackage{color}
\usepackage{graphicx}
\usepackage{placeins}
\usepackage[utf8]{inputenc} 
\usepackage[T1]{fontenc}
\usepackage[colorlinks=true,linktoc=page,linkcolor=magenta,citecolor=cyan]{hyperref}
\usepackage{chngcntr}

\newcommand\bea{\begin{eqnarray}}
\newcommand\eea{\end{eqnarray}}
\newcommand\beq{\begin{equation}}  
\newcommand\eeq{\end{equation}}

\newcommand{\non}{\nonumber}  

\begin{document}
\textheight=23.8cm

\title{Particle dynamics and ergodicity-breaking in twisted-bilayer optical lattices}
\author{Ganesh C. Paul}
\affiliation{Institut f\"{u}r Mathematische Physik, Technische Universit\"{a}t Braunschweig, D-38106 Braunschweig, Germany}
\author{Patrik Recher}
\affiliation{Institut f\"{u}r Mathematische Physik, Technische Universit\"{a}t Braunschweig, D-38106 Braunschweig, Germany}
\affiliation{Laboratory for Emerging Nanometrology Braunschweig, D-38106 Braunschweig, Germany}
\author{Luis Santos}
\affiliation{Institut f\"{u}r Theoretische Physik, Leibniz Universit\"{a}t, 30167 Hannover, Germany}
\date{\today}

\begin{abstract}

Recent experiments have realized a twisted bilayer-like optical potential for ultra-cold atoms, which in contrast 
to solid-state set ups may allow for an arbitrary ratio between the inter- and intra-layer couplings. For commensurate Moir\'e twistings a large-enough inter-layer coupling results in particle transport dominated by 
channel formation. For incommensurate twistings, the interlayer coupling acts as an effective disorder strength. Whereas for weak couplings the whole spectrum remains ergodic, at a critical value part of the eigen-spectrum transitions into multifractal states. A similar transition may be observed as well as a function of an energy bias between the two layers. Our theoretical study reveals atoms in a twisted-bilayer system of square optical lattices as an interesting new platform for the study of ergodicity breaking and multifractality.

\end{abstract} 
\maketitle


\section{Introduction}

Twisted bilayer graphene~\cite{bistritzer2011,dos2007,andrei2020,balents2020,torma2022} has attracted broad attention owing to the observation of unconventional superconducting~\cite{cao2018a,yankowitz2019,oh2021,yankowitz2019tuning} and correlated insulating behaviour~\cite{cao2018b,codecido2019,nuckolls2020,cao2020tunable}. A small rotation of one of the layers leads to the vanishing of Fermi velocity around the Dirac point giving rise to an almost flat band~\cite{dos2007,trambly2010,bistritzer2011,lisi2021observation}. 
These quasi-flat bands, ideal to observe strongly correlated phenomena, are obtained only for very small twist angles $\lesssim  1^\circ$ in solid state systems, as the inter-layer coupling is much smaller than the intra-layer one~\cite{morell2010,tarnopolsky2019,lisi2021}.
 
Ultra cold gases in optical potentials may provide an interesting highly-controllable 
platform for the study of the physics of twisted-bilayer lattices. These systems
allow for a basically arbitrary ratio between inter-layer and intra-layer couplings. In addition, ultra-cold 
gases are in principle defect-free, although suitable impurities can be added in a controllable way, making this platform ideal for understanding the effects of disorder. Different proposals have been recently put forward ~\cite{gonzalez2019,salamon2020,luo2021spin,lee2022} to simulate twisted-bilayer-like potentials using ultracold atoms. 
In particular, Ref.~\cite{gonzalez2019} has proposed the use of two internal states that (in a synthetic dimension) play the role of the two layers. The twisted lattices result from a state-dependent optical potential, such that one state experiences an optical lattice tilted at an angle from the lattice experienced by the other state. A microwave or two-photon Raman coupling induces an effective inter-layer hopping. An alternative proposal was introduced in Ref.~\cite{salamon2020}, also using a synthetic dimension for the bilayer geometry, with lattices without twisting but with a spatially-dependent inter-layer hopping. Very recently, 
similar ideas as those of Ref.~\cite{gonzalez2019}, have been employed 
to realize experimentally for the first time a twisted-bilayer optical potential~\cite{meng2023atomic}, in which a Bose-Einstein condensate was loaded, opening an interesting novel platform for the study of superfluids in twisted-bilayer lattices. 

In recent years, atoms in optical lattices have been shown to provide a suitable platform to study experimentally both single- and many-body localization. 
In particular, the use of bichromatic lattices has allowed for the realization of the one-dimensional quasi-disordered (Aubry-Andr\'e) model~\cite{aubry1980, Roati2008, Schreiber2015}~, characterized by a phase transition at a critical quasi-disorder strength between a fully localized and a fully ergodic eigenstate spectrum.  Recent experiments have also realized two-dimensional optical quasi-crystals~\cite{Viebahn2019}. An alternative, also highly controllable set-up for the study of localization, is provided by photonic lattices, in which a spatial direction plays the role of an effective time dimension. Interestingly, in addition to one-dimensional geometries~\cite{lahini2009}, very recent 
experiments have analyzed the wave dynamics in two-dimensional (monolayer) photonic moir\'e-like lattices~\cite{wang2020}, revealing a localization-to-delocalization transition~\cite{huang2016}. 

The above-mentioned realization of twisted-bilayer-like optical lattices opens intriguing questions concerning particle dynamics and ergodicity breaking in these potentials, which we theoretically address in this paper for the case of coupled square lattices.
Whereas in solid-state set-ups the interlayer coupling is very small compared to the intra-layer one~(typically $5$ to $10$ times smaller), it may be potentially dominant in optical lattice platforms, resulting for commensurate twistings in particle transport dominated by the formation of channels. 
Moreover, for incommensurate twistings and due to the finite spatial range of the inter-layer coupling, this coupling acts as an effective quasi-disorder strength. Whereas below a given coupling threshold the whole eigenspectrum remains ergodic, at a critical coupling part of the spectrum experiences a transition into non-ergodic extended (multi-fractal) states. A similar transition may be observed, alternatively, by employing an energy bias between the two layers. Our results show hence that the combination of moderately strong interlayer coupling and incommensurate twist angles 
makes twisted-bilayer optical lattices a new suitable platform for the study of multifractality. \\
 
The remainder of the paper is organized as follows. In Sec.~\ref{sec:Twisted}, we describe the twisted-bilayer model. Section~\ref{sec:Moire} is devoted to the particle dynamics in the case of commensurate twistings, while the dynamics in incommensurately twisted bilayers is presented in Sec.~\ref{sec:Incommensurate}. The impact of inter-layer bias is discussed in Sec.~\ref{sec:Bias}. Finally, we summarize our results in Sec.~\ref{sec:Conclusions}.
 

\section{Optical twisted bilayers}
\label{sec:Twisted}

\begin{figure}[t!]
	\centering
	\includegraphics[width=0.3\textwidth, clip=true]{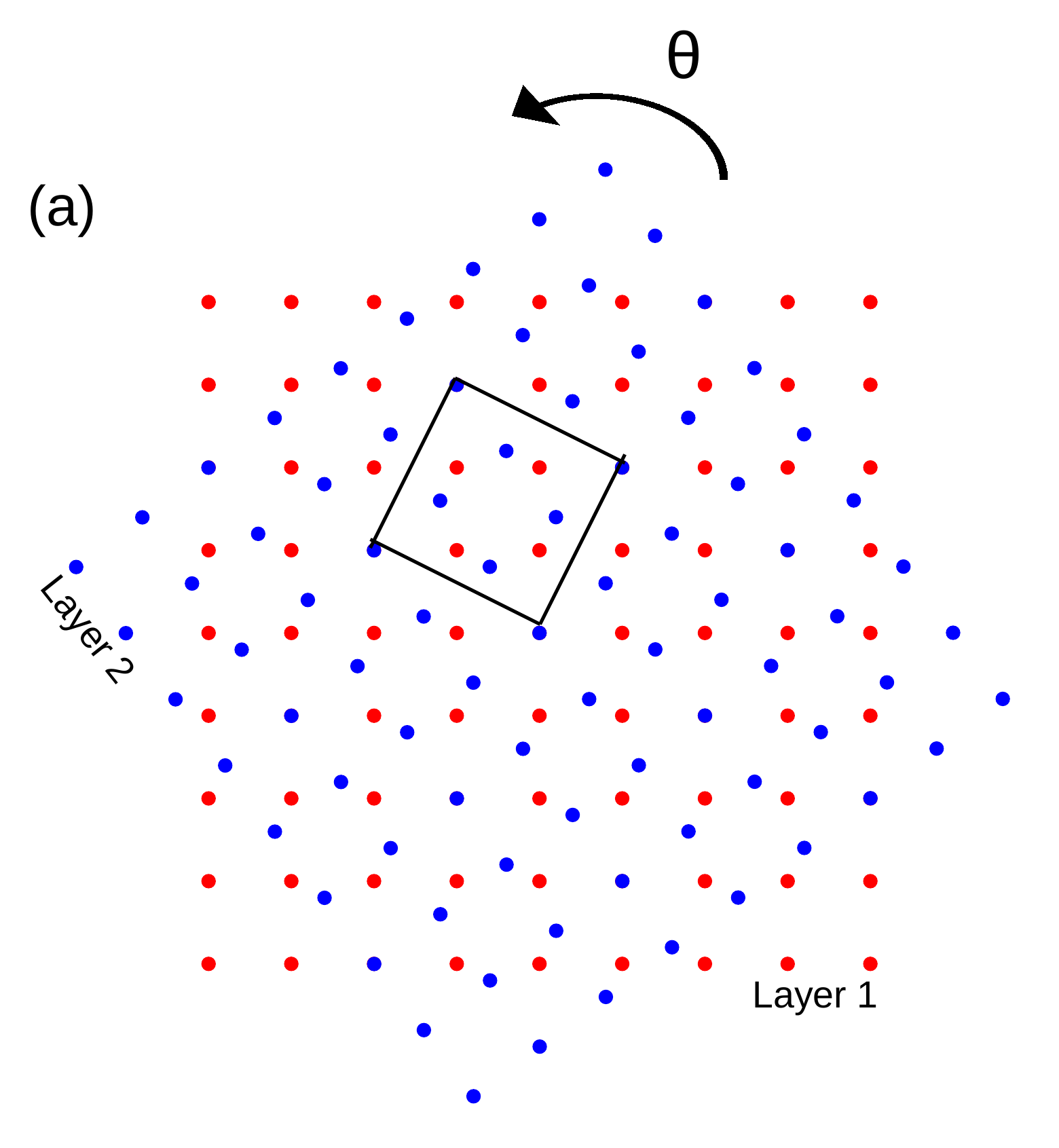}\hfill
	\includegraphics[width=0.4\textwidth, clip=true]{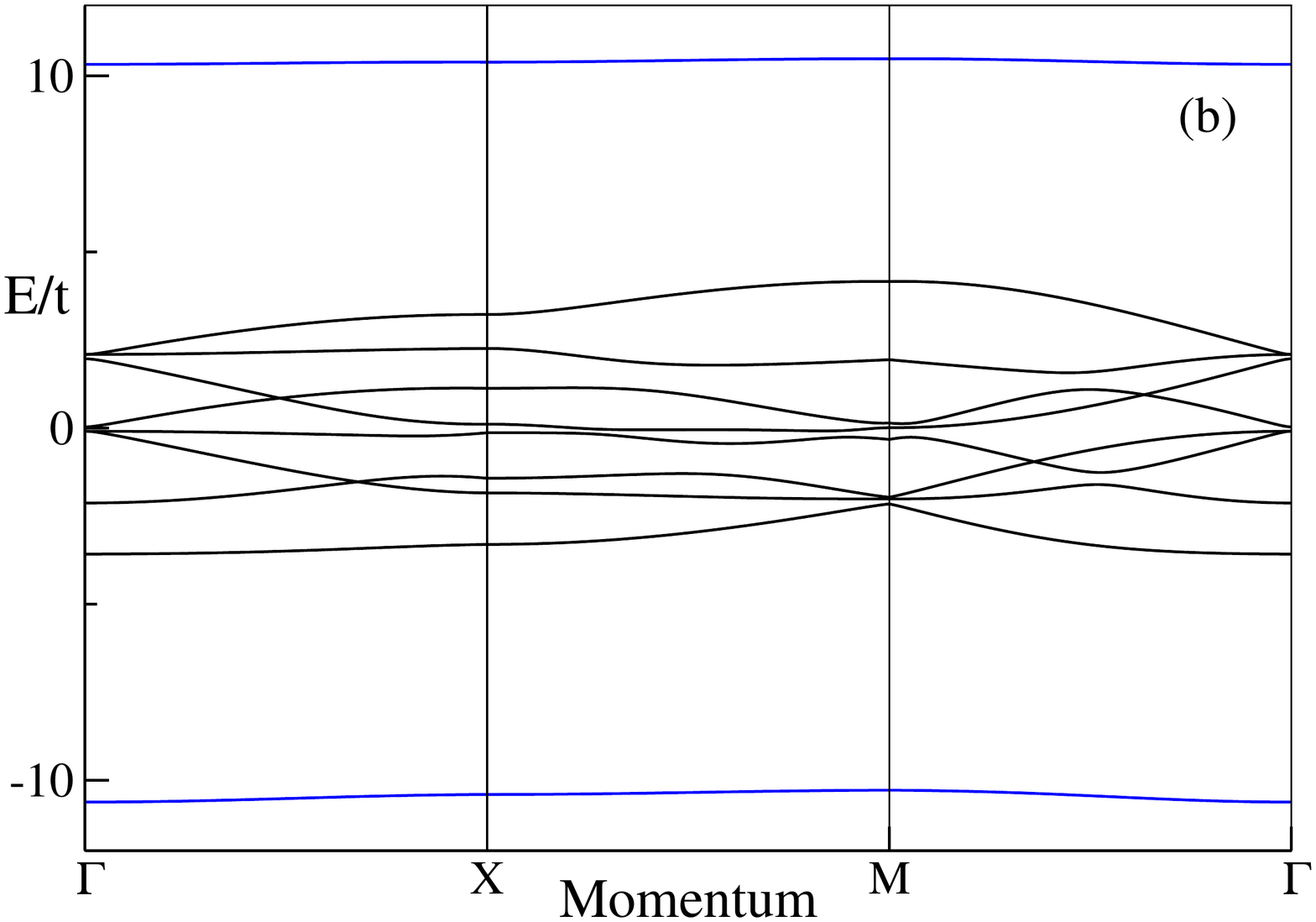}
	\caption{(a) Twisted-bilayer structure for a twisting angle $\theta=\theta(2,1)$. Red~(blue) sites correspond to layer $1$~($2$). The elementary cell is indicated by a black square. (b) Band structure along the high-symmetry line $\Gamma-X-M-\Gamma$ for the same lattice with $t_\perp/t=10$, $l_0/a=0.15$, and $\Delta=0$. The quasi-flat bands are indicated in blue.}
	\label{band}
\end{figure}

In the following, we consider two layers of square optical lattices~(see Fig.~\ref{band}(a)), where one layer is twisted by an angle $\theta$ with respect to the other. A possible way of implementing 
such an optical potential, recently proposed in Ref.~\cite{gonzalez2019} and realized experimentally in Ref.~\cite{meng2023atomic}, employs an atom in two different internal states. In this scenario, which we assume below, the bilayer structure is provided by the synthetic dimension given by the two internal states, whereas a state-dependent potential results in the twisted-bilayer geometry. For a square lattice, a moir\'e pattern is achieved  
for $\theta=\theta(m,n)$\, where $\theta(m,n)=\arccos(\frac{2mn}{m^2+n^2})$ ($n,m\in\mathbb{Z}$). 



We are interested in the dynamics of a single particle, and hence we do not account for interaction terms. The results should remain, however, valid as long as the lattice filling is sparse-enough. The system is characterized by the Hamiltonian:
\bea
H&=&- \Delta \sum_{\alpha,j}  (-1)^{\alpha} |\alpha,j\rangle\langle \alpha,j|-t\sum_{\alpha=1,2}\sum_{\langle j,j'\rangle}
|\alpha,j\rangle\langle \alpha,j'|\non \\
&-&\sum_{jj'}t_\perp (j,j') \left[ |1,j\rangle\langle 2,j'| + \mathrm{H. c.} \right ], 
\label{Ham}
\eea
where $|\alpha,j\rangle$ is the state in which the particle is in layer $\alpha$ in site $j=(j_x,j_y)$. The site in layer $1$~($2$) 
is located at the position $\vec R_{1,j}=j_x \vec e_x + j_y \vec e_y$~($\vec R_{2,j}=j_x \vec e_{x'} + j_y \vec e_{y'}$), with 
$\vec e_{x'}=\cos\theta\vec e_x + \sin\theta\vec e_y$, 
$\vec e_{y'}=-\sin\theta\vec e_x + \cos\theta\vec e_y$. The first term of Eq.~\eqref{Ham} denotes an uniform bias between the layers, characterized by the bias strength $\Delta$, which in the synthetic dimension scenario amounts for a level-dependent shift (e.g. using a magnetic field or an optically-imposed Stark-shift). The intra-layer hopping, given by the rate $t$, occurs only to nearest neighbors, denoted by $\langle j,j' \rangle$. 

The rate $t_\perp(j,j')$ characterizes the hopping between a site $j$ in layer $1$ and site $j'$ in layer $2$. In the considered scenario, such a coupling 
occurs between different internal states, and it is given either by a microwave or two-photon optical Raman coupling. In stark contrast to regular cubic lattices, the sites in both layers are generally not on top of each other. For a sufficiently strong lattice, we may approximate the on-site Wannier functions in each layer as a Gaussian of with $l_0=\frac{a}{\pi s^{1/4}}$, with $a$ the lattice spacing characterizing the square lattice in both layers, and $s$ the lattice depth in recoil units. As a result, the inter-layer hopping acquires the form~\cite{gonzalez2019}:  
\bea
t_\perp(j,j')= t_\perp e^{-|\vec R_{1,j}-\vec R_{2,j'}|^2/(4l_0^2)}. 
\label{eq:tp}
\eea
Reference~\cite{gonzalez2019} considered for simplicity the case of $l_0=0$, in which only sites exactly on top of each other may undergo inter-layer coupling with a rate $t_\perp$. We show below that the finite Gaussian width $l_0$ plays a crucial role in the actual particle dynamics in the bilayer-like optical potential. Note as well, that in contrast to solid-state scenarios where typically $t_\perp/t\ll 1$ and hardly tunable, in the optical lattice platform $t_\perp$ is easily tunable and may be much larger than the intra-layer hopping $t$. As shown below, this opens interesting possibilities for the dynamics for both commensurate and incommensurate twisting angles.

In the following, we consider for simplicity, unless otherwise indicated, a twist angle $\theta$ in the vicinity of the magic angle $\theta(2,1)=36.87^\circ$, although our conclusions are general for the dynamics in the vicinity of any commensurate twist angle. 
The choice of $\theta(2,1)$ is justified by the small number of sites~($5$ in each layer) per Moir\'e unit cell, which greatly simplifies the analysis of the system, compared to solid-state platforms, where due to the very small twisting angle, a Moir\'e cell may contain tens of thousands of sites~\cite{trambly2010localization}. For $\theta=\theta(2,1)$, only $2$ of the $10$ sites are on top of each other~(directly-connected sites), see Fig.~\ref{band}(a).




\begin{figure*}[t]
	\centering
	\includegraphics[width=2\columnwidth, clip=true]{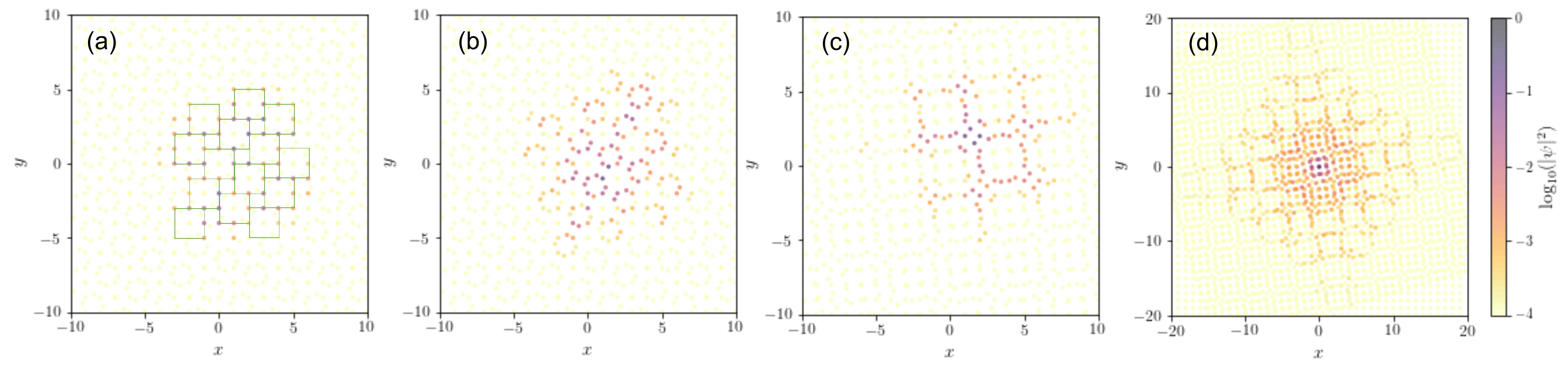}\hfill
	\caption{Commensurate twisting. Panels (a) and (b) show the probability density at time $\tau=2/t$ for $\theta=\theta(2,1)$, $t_\perp/t=100$, and $l_0=0$ and $l_0/a=0.15$, respectively, for a particle initially at site $j=(1,0)$ of layer $1$. In panel~(a), green lines indicate the effective lattice with out-projected sites mentioned in the text. Panel (c) depicts the probability density at $\tau=6/t$ for $\theta=\theta(4,3)$ for $l_0/a=0.15$, and $t_\perp=50t$, for a particle initially in $j=(2,2)$ of layer $1$. Panel (d)
	shows, at $\tau=10/t$, the particles distribution for an initial Gaussian wavepacket in layer $1$, $\frac{1}{\sqrt{2\pi\sigma^2}} e^{-|\vec{R}_{1,j}-\vec{R}_{1,j_0}|^2/2\sigma^2}$,
	of width $\sigma/a=1$, centered at site $j_0=(0,0)$, in a lattice with $t_\perp=10 t$, and $\theta(4,3)$. In all the figures, the $x$ and $y$ axes are in units of the lattice spacing $a$.}
	\label{fig:fig2new}
\end{figure*}


\section{Particle dynamics in Moir\'e lattices}
\label{sec:Moire}

Figure~\ref{band}(b) shows the corresponding band structure for $\theta=\theta(2,1)$, $\Delta=0$ and $t_\perp=10 t$, along the symmetry line $\Gamma-X-M-\Gamma$,  with $\Gamma=(0,0)$, $X=\frac{2\pi}{5a}(\frac{1}{2},1)$, and $M=\frac{2\pi}{5a}(-\frac{1}{2},\frac{3}{2})$. As shown in Fig.~\ref{band}(b), when $t_\perp/t\gg 1$, the spectrum presents an uppest and a lowest band~(in blue), with an energy $\simeq \pm t_\perp$, which originate from directly-connected sites. The large inter-layer coupling projects out those sites, which build a separate grid. Since directly-connected sites are far apart, hopping between them only occurs in high-order in perturbation theory, resulting in almost flat bands. Particles initially placed in those sites remain hence basically localized.

The rest of the not-directly connected sites form separated bands around zero energy. For $l_0=0$, they cannot participate directly in the inter-layer hopping. For $t_\perp\to\infty$, a particle starting in one of those sites in the upper~(lower) layer would be disconnected from the lower~(upper) layer. The particle would experience an effective square lattice with out-projected sites~(the directly-connected ones), characterized by four sites in the elementary cell~(see Fig.~\ref{fig:fig2new}(a)). As a result, the band spectrum would present two degenerate sets of four bands. This degeneracy is lifted, even for $l_0=0$, at finite $t_\perp/t\gg 1$ due to processes of order ${\cal O}(t^2/t_\perp)$ induced by virtual couplings between directly- and non-directly-connected sites. 

A single particle in the twisted optical lattice is 
described by the state
$\sum_{\alpha,j} c_{\alpha j}(t)\, |\alpha, j\rangle
$, with the probability amplitudes given by the Schr\"odinger equations:
\bea
i\hbar \frac{\partial{c_{\alpha,j}}}{\partial{\tau}}&=&-t\sum_{\langle jj'\rangle} c_{\alpha,j'}-\sum_{j'} t_\perp (j,j') c_{\bar\alpha,j'},
\label{eq:Schroedinger}
\eea
with $\bar\alpha=2$~($1$) for $\alpha=1$~($2$). We solved these equations by standard Runge-Kutta methods, using absorptive boundary conditions, which allow for the analysis of the particle expansion and eventual localization without the need of very large lattices. 

We consider the evolution of a particle initially placed in a not-directly connected site. As mentioned above, if $l_0=0$, 
a sufficiently large $t_\perp/t$ results in the motion of the particle in an effective square lattice with out-projected sites, see Fig.~\ref{fig:fig2new}(a) for $t_\perp/t=50$. The situation is radically different for a finite $l_0=0.15 a$, corresponding to a lattice depth $s=20$. For $\theta=\theta(2,1)$, due to the particularly simple elementary cell,
particles move amongst all the non-directly connected sites of both layers~(see Fig.~\ref{fig:fig2new}(b)), and the expansion dynamics is independent of the chosen initial not directly-connected site. For smaller Moir\'e angles, the dynamics for $t_\perp/t\gg 1$ strongly depends on the initial site, because the central bands break into separate sub-bands characterized by very different transport properties. Whereas some sites form quasi-isolated islands, other sites connect efficiently to a net of sites building transport channels~(see Fig.~\ref{fig:fig2new}(c)). 
Hence, for finite times, of practical relevance in typical experiments, the formation of channels dominates particle transport in the moir\'e lattice. As a result, a particle initially distributed amongst various sites, generally undergoes a bimodal expansion dynamics, characterized by partial channel-like expansion and partial quasi-localization in poorly-connected sites, 
see Fig.~\ref{fig:fig2new}~(d).


\section{Dynamics in incommensurate bilayers}
\label{sec:Incommensurate}

As shown above, for general $\theta=\theta(m,n)$, the particle 
dynamics is characterized by the splitting of the spectrum into separate bands, and the corresponding formation for large 
$t_\perp/t$ of quasi-isolated regions and effective lattice channels that dominate the (partial) particle expansion. This picture is distorted when the tilting angle departs from commensurability. We introduce at this point the departure angle $\phi$, such that $\theta=\theta(m,n)+\phi$. This incommensurability, together with the finite Gaussian range $l_0$, results in a spatial quasi-disorder of the inter-layer coupling, which may severely affect the particle dynamics, resulting in ergodicity breaking and eventually localization. 
In this section, we analyze in detail 
the role played by the inter-layer coupling, whereas  Sec.~\ref{sec:Bias} is devoted to the effect of the bias $\Delta$.

 
 
\begin{figure*}[t!]
\includegraphics[width=2\columnwidth, clip=true]{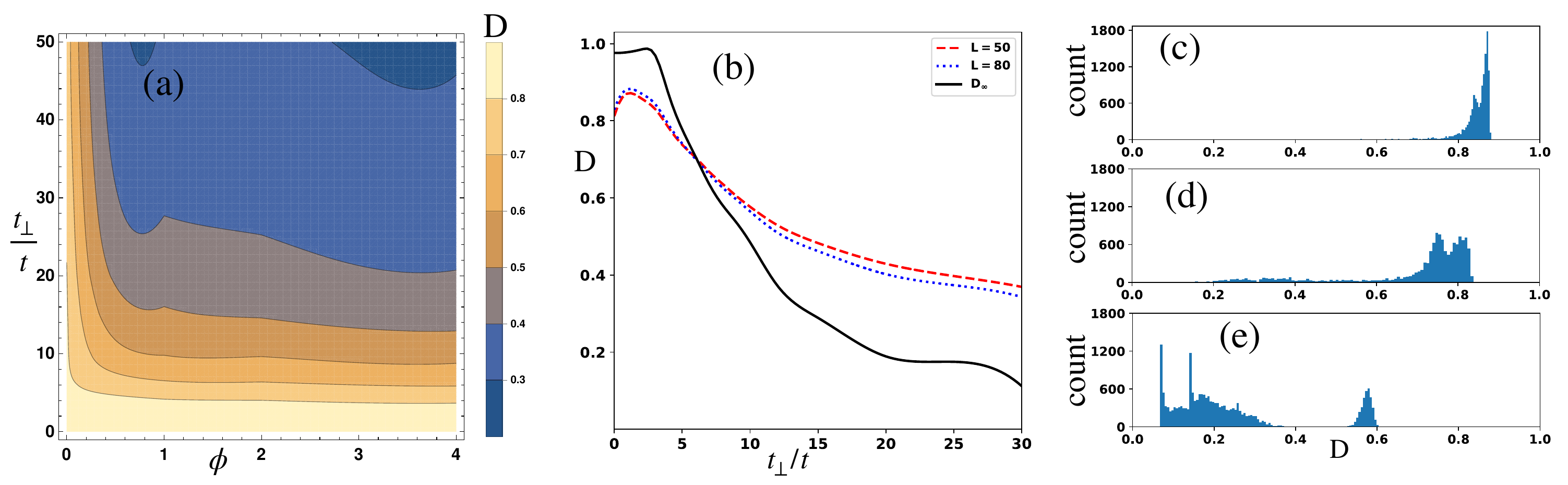} \hfill
\caption{Eigenstate properties for incommensurate twisting angles.  (a) Averaged fractal dimension $D(N)$ for $N=80\times 80$ sites, as a function of $t_\perp/t$ and $\phi$. (b) $D(N)$ for $\phi=3^\circ$, and a lattice with $N=50\times 50$~(red dashed) and $80\times 80$~(blue dotted) sites. The black solid curve depicts $D_\infty$~(see text). (c-e) Histogram of the distribution of $D_2(\psi)$ values for $t_\perp/t=3$~(c), 
$12$~(d) and $100$~(e). In all the figures we consider $\theta=\theta(2,1)+\phi$, $l_0/a=0.15$ and $\Delta=0$.}
\label{fig:fig4_new}
\end{figure*}


\subsection{Eigenstates}

We first focus on the localization properties of the lattice eigenstates, $|\psi\rangle = \sum_{\alpha,j} c_{\alpha,j}(\psi) |\alpha,j\rangle$, which are well characterized by the moments 
 \beq
 I_q (\psi)\,=\,\sum_{\alpha,j} |c_{\alpha,j}(\psi)|^{2q}\,\propto\, N^{-D_q(\psi)(q-1)},
 \eeq
where $N$ is the total number of sites in each layer, and $D_q(\psi)$ are the fractal dimensions. In particular, the inverse participation ratio~(IPR)~\cite{wegner1980,evers2000,murphy2011},  $I_2(\psi)$, is given by the inverse of the number of sites in which the eigenstate $|\psi\rangle$ has a significant support. Localized~(ergodic) states are characterized by $D_q(\psi)=0$~($D_q(\psi)=1$), whereas intermediate $q$-dependent  $0<D_q(\psi)<1$ implies an extended but non-ergodic character, and a multifractal geometry~\cite{jagannathan2021fibonacci,pook1991,chamon1996,deng2016,mace2019,deng2019one,sarkar2022,reisner2022experimental,de2014anderson}. 

In the following, we focus our analysis on the IPR, 
determining for a lattice with $N$ sites, up to $80\times 80$, and open boundary conditions, 
the fractal dimension of each eigenstate 
$D_2(N;\psi)=\log I_2(\psi) / \log (N)$.
In order to provide a global characterization 
of the system, we evaluate the averaged value of 
the fractal dimension over all eigenstates, $D(N)=\frac{1}{N}\sum_\psi D_2(N; \psi)$. 

Figure~\ref{fig:fig4_new}(a) shows $D(N)$ for $N=80\times 80$ sites, as a function of the interlayer hopping $t_\perp/t$ and the departure angle $\phi$~(note that we limit to $0<\phi<4^\circ$ in order to avoid nearby commensurate tilting angles; the results for $\phi<0$ are very similar). For $t_\perp=0$, each layer constitutes an independent disorder-free square lattice, characterized by ergodic eigenstates, and correspondingly band expansion. The eigenstates are also strictly speaking ergodic for a Moir\'e lattice~($\phi=0$) irrespective of the value of $t_\perp/t$. However, as mentioned above, the corresponding bands may be significantly flat, leading to potentially very long time scales for the band-expansion dynamics. Note as well, that the formation of separated bands characteristic of 
large-enough $t_\perp/t$ values reduces the lattice support of the eigenstates, resulting for large $t_\perp/t$ in a finite-size-induced deviation from the expected value $D=1$. We discuss this finite-size effects below.

For a non-zero tilting deviation $\phi$, a finite 
$t_\perp$ results in an effective two-dimensional quasi-disordered spatial dependence of the 
inter-layer hopping amplitude. Note that this is so, crucially, because $l_0>0$. A vanishing $l_0/a$ would result for a finite $\phi$ in the almost complete decoupling of the layers, and hence on ergodic, basically monolayer, dynamics. As a result, $t_\perp/t$ would not play the role of quasi-disordered strength discussed below. Note as well, that the quasi-disorder is in principle established even for very small angle deviations. However,  
a lattice with a very small $\phi<1^\circ$ is barely distinguishable from a Moir\'e lattice for the system sizes considered in our numerics (and for typical experimental sizes), resulting in the enhancement of the value of $D$ observed in Fig.~\ref{fig:fig4_new}(a). For large $\phi>1^\circ$ values, the results are approximately $\phi$ independent.
As seen in Fig.~\ref{fig:fig4_new}(a), the ergodic character of the whole eigenspectrum is maintained at low-enough inter-layer couplings, with $D(N)>0.85$ for 
$t_\perp/t<4$. Beyond that value, $D$ decays markedly
reaching values $D(N)\lesssim 0.3$ already for $t_\perp/t\simeq 25$.

Finite-size effects pose a major difficulty when studying the localization properties and in particular 
the fractal dimension. These effects may be to a large extent mitigated using the following argument.
Note that for a given eigenstate $|\psi\rangle$ in a system with $N$ sites, 
$I_2(\psi)=\gamma(\psi) / N^{D_2(\psi)}$. 
Hence the evaluated fractal dimension 
$D_2(N;\psi) = D_2(\psi) + \frac{\log\gamma(\psi)}{\log N}$. 
Assuming that the deviation averaged over all eigenstates $\frac{1}{N}\sum_\psi\log\gamma(\psi)$ is approximately $N$ independent, we may then employ the following ansatz for 
the relation between the averaged fractal dimension  $D(N)$ for the case of $N$ sites 
and the value $D_\infty$ expected for 
an infinitely large system:
\begin{equation}
D(N)=D_\infty+\alpha/log(N), 
\end{equation}
where $D_\infty$ and $\alpha$ are determined by fitting 
our results for different system sizes, up to $80\times 80$ sites. Figure~\ref{fig:fig4_new}(b) shows our results for $D_\infty$ 
for $t_\perp/t=10$, $l_0/a=0.15$, and a deviation $\phi=3^\circ$ from the Moir\'e angle $\theta(2,1)$. 

The extrapolated results confirm the qualitative picture observed in Fig.~\ref{fig:fig4_new}(a). The spectrum shows a clear change of character at $t_\perp/t\simeq 3$. 
For weaker inter-layer coupling $D_\infty\simeq 1$, and hence the whole spectrum is ergodic. Figure~\ref{fig:fig4_new}(c) shows the distribution of $D_2(N;\psi)$ for $N=80\times 80$ sites, with the expected peak at large $D$ values~(only limited by finite size effects).
At $t_\perp/t\simeq 3$, whereas part of the spectrum remains ergodic, the rest undergoes an ergodic-to-non-ergodic transition, resulting in intermediate $D_\infty$ values.
These intermediate values are not a finite-size effect. Note in this sense, that 
for $t_\perp/t\simeq 6$, the $D(N)$ curves with different $N$ cross, indicating that around that value the spectrum is $N$ independent~($\alpha\simeq 0$ in the expression above). Note as well that the non-ergodic eigenstates have not a localized, but rather an extended nature. This is evident from Fig.~\ref{fig:fig4_new}(d), where we depict the distribution of $D_2(N;\psi)$ for $t_\perp/t=12$. This distribution shows in addition to basically ergodic states, a large number of states well within the intermediate regime of $D_2$ values.
Further increasing $t_\perp/t$~(see Fig.~\ref{fig:fig4_new}(e) for $t_\perp/t=100$) results in a the displacement of the bulk of the spectrum to low $D_2$ indicating localization (although 
part of the states remain with a clear non-ergodic extended character even for such a strong inter-layer coupling). While the behavior of D around the twist angle $\theta(2,1)$ is presented in Figure~\ref{fig:fig4_new}(a), we have verified that the physics is very similar for other twist angles.  


 \subsection{Expansion dynamics}

\begin{figure}[b!]
\includegraphics[width=0.35\textwidth, clip=true]{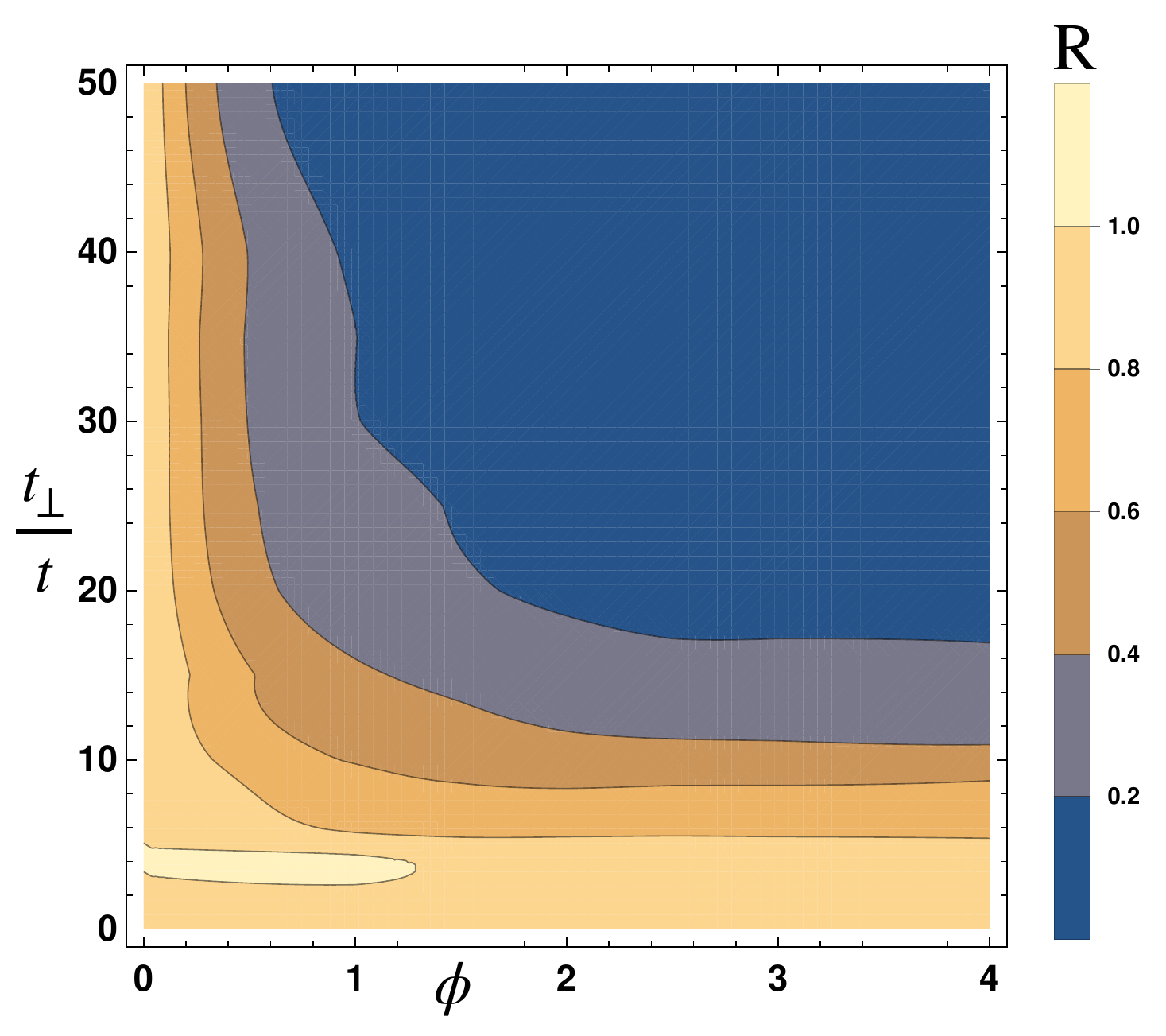}
\caption{Expansion dynamics for incommensurate twist angles.
Expansion radius $R(\tau=16/t)$ as a function of $\phi$ and $t_\perp/t$ for the same case of Fig.~\ref{fig:fig4_new}(a).
}
\label{fig:fig5_new}
\end{figure}

The change in the nature of the eigenstates when increasing $t_\perp/t$ and $\phi$ translates into a marked modification of the expansion dynamics of an initially localized wavepacket~(at $\tau=0$). We characterize the particle expansion at a given time $\tau>0$ by means of the average distance $\overline{r}$, from the initial central position $\vec{R}_{in}$ of the particle wavepacket: 
\beq
\overline{r}(\tau)^2= \sum_{\alpha=1,2} \sum_j |\vec{R}_{\alpha, j}-\vec{R}_{in}|^2\, |c_{\alpha,j}(\tau)|^2. 
\eeq
In order to assess the effect of twisting incommensurability
we compare this radius with the one, $\overline{r}_0(\tau)$,  expected for $\phi=0$, defining $R(\tau)=\overline{r}(\tau)/\overline{r}_0(\tau)$. This normalization is necessary, since, as mentioned above, the dynamics in a Moir\'e lattice may slow down very significantly with $t_\perp/t$ due to the appearance of quasi-flat bands. Figure~\ref{fig:fig5_new} shows for different values of $\phi$ and $t_\perp/t$, the normalized radius $R(\tau=16/t)$, for an initial wavepacket centered at $(0,0)$ with width $\sigma/a=1$. The qualitative behavior of $R$ mirrors that 
of the fractal dimension $D$ in Fig.~\ref{fig:fig4_new}(a). For $t_\perp/t<4$ the expansion dynamics is basically 
identical to that of the commensurate case. In the vicinity of $\phi=0$, we 
observe again that due to finite-size~(and also finite-time) limitations, a small $\phi$ is almost indistinguishable from a Moiré bilayer. In contrast, the results are only weakly dependent on $\phi$ for $\phi>1^\circ$. 
In that regime, $R(\tau)$ decreases very markedly with growing $t_\perp/t$, indicating the onset of non-ergodic dynamics, and eventually localization.

 
 
\begin{figure*}[t!]
\includegraphics[width=2\columnwidth, clip=true]{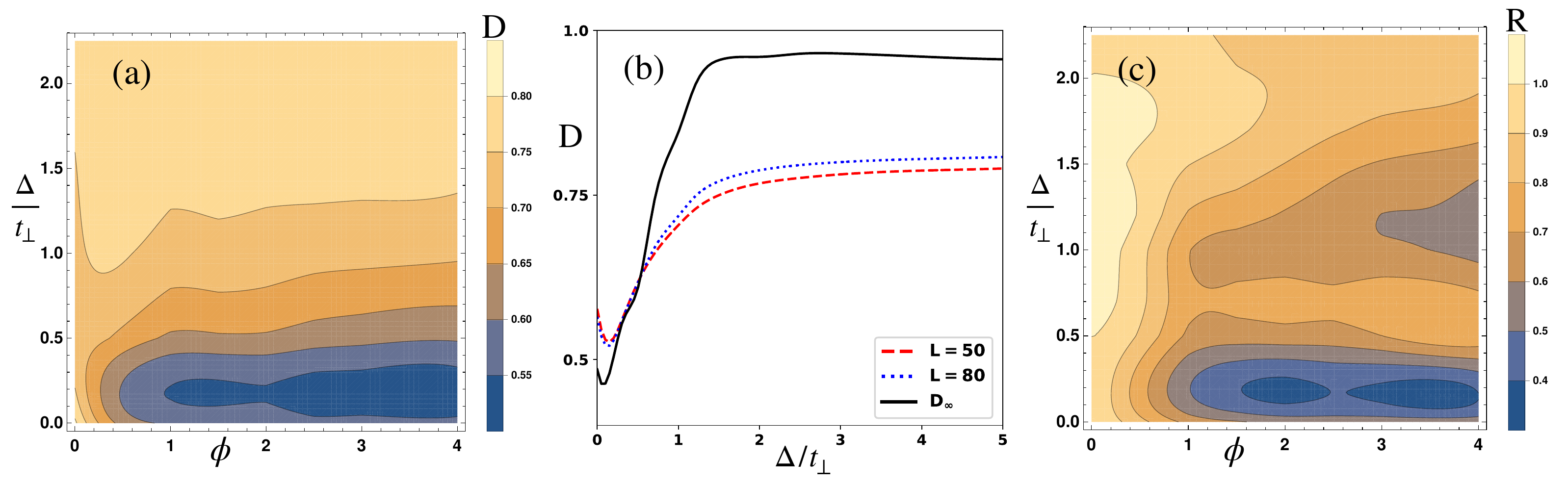}
\caption{Effect of the inter-layer bias.
(a) Averaged fractal dimension $D(N)$ for $N=80\times 80$ sites as a function of $\Delta/t_\perp$ and $\phi$. 
(b) $D(N)$ for $\phi=3^\circ$, and a lattice with $N=50\times 50$~(red dashed) and $80\times 80$~(blue dotted) sites. The black solid curve depicts $D_\infty$~(see text).
(c) Expansion radius $R(\tau=20/t)$ as a function of $\phi$ and $t_\perp/t$ for the same case of Fig.~(a).
In all the figures, $t_\perp/t=10$, $l_0=0.15a$, and  $\theta(2,1)+\phi$.}
\label{fig:fig6_new}
\end{figure*}


3\section{Inter-layer bias}
\label{sec:Bias}

The previous section has shown that the inter-layer coupling acts as an effective quasi-disorder strength 
that controls the degree of ergodicity-breaking in the system. As discussed in this section, a similar role may be played as well by the bias $\Delta$, i.e. by the energy off-set between the two layers in Eq.~\eqref{Ham}. Figure~\ref{fig:fig6_new}~(a) depicts the averaged fractal dimension $D$ as a function of $\phi$ and $\Delta/t_\perp$, for an $80\times 80$ lattice, with $\theta=\theta(2,1)+\phi$, $l_0/a=0.15$ and $t_\perp/t=10$.  

As in the previous section, the region of small $\phi$ presents an anomalously large $D$ due to the finite-size quasi-indistinguishability from the Moiré bilayer case. Increasing the bias $\Delta$ renders the inter-layer hopping off-resonant, eventually resulting for a large-enough $\Delta/t_\perp$ in an effective decoupling of the layers. Since a decoupled layer is a regular square lattice, a large-enough bias leads to the retrieval of ergodicity irrespective of $\phi$. The extrapolated value $D_\infty$~(see Fig. ~\ref{fig:fig6_new}(b)) shows that the whole spectrum remains ergodic 
for $\Delta/t_\perp>1.5$. Below that value, part of the spectrum becomes 
multifractal. Note that also in this curve we may identify a point, at $\Delta/t_\perp\simeq 0.5$ at which the curves $D(N)$ for different $N$ cross.

As in the previous section, the time dependence of the normalized averaged distance, $R$, mirrors the spectral properties~. Similar to the previous section, we normalize $R$ to the value expected for $\phi=0$. Figure~\ref{fig:fig6_new}~(c) depicts $R(\tau=20/t)$ as a function of $\phi$ and $\Delta/t$, showing a 
good qualitative agreement with Fig.~\ref{fig:fig6_new}(a). The average distance is almost independent of $\phi$ for large bias implying an extended phase in the effective single-layer regime. A smaller bias leads to a markedly non-ergodic dynamics. Interestingly, the largest deviation from ergodicity occurs for $\phi>1^\circ$ not at $\Delta=0$, but rather at a finite $\Delta/t\simeq 2$.


\section{Conclusions} 
\label{sec:Conclusions}
Particle dynamics in a twisted bilayer optical lattice 
presents a non-trivial dependence on the lattice depth (which controls the hopping $t$ and the width $l_0$ of the inter-layer Gaussian coupling), the inter-layer coupling strength $t_\perp/t$, the inter-layer bias $\Delta/t$, and the tilting angle $\theta$. Crucially, 
in contrast to solid-state twisted bilayer set ups, the inter-layer coupling 
can be controlled basically at will, and 
may be much larger than the intra-layer one. 
As a result, the inter-layer coupling 
may play a crucial role in the particle dynamics. For commensurate Moir\'e-like lattices, the eigenstates are in any case ergodic, but due to the flatness of the Moir\'e bands, for relevant experimental time scales transport for $t_\perp/t\gg 1$ is dominated  by channel formation. For incommensurate lattices, and due to the finite range of the inter-layer coupling, the coupling rate $t_\perp$ acts as an effective quasi-disorder strength. Whereas the spectrum remains fully ergodic for small $t_\perp/t\lesssim 3$~(for $\Delta=0$), a larger inter-layer coupling induces the transition of part of the spectrum into non-ergodic (but still extended) states. Similarly, ergodicity is recovered for a sufficiently large inter-layer bias, whereas reducing the bias induces again a partial ergodic-to-non-ergodic transition.
The spectral properties may be revealed 
from the analysis of the expansion dynamics of an initially localized wavepacket. Our results hence show that twisted-bilayer optical lattice set ups provide an interesting controllable platform for the study of multifractality. 
Furthermore, multi-layer set-ups may allow for the study of 
particle transport in the 2D-to-3D cross-dimensional regime, an intriguing possibility for future studies. 

\acknowledgments{We acknowledge support of the Deutsche Forschungsgemeinschaft (DFG, German Research Foundation) under Germany's Excellence Strategy -- EXC-2123 Quantum-Frontiers -- 390837967.}

\bibliography{TBG_sq} 

\begin{thebibliography}{43}%
\makeatletter
\providecommand \@ifxundefined [1]{%
 \@ifx{#1\undefined}
}%
\providecommand \@ifnum [1]{%
 \ifnum #1\expandafter \@firstoftwo
 \else \expandafter \@secondoftwo
 \fi
}%
\providecommand \@ifx [1]{%
 \ifx #1\expandafter \@firstoftwo
 \else \expandafter \@secondoftwo
 \fi
}%
\providecommand \natexlab [1]{#1}%
\providecommand \enquote  [1]{``#1''}%
\providecommand \bibnamefont  [1]{#1}%
\providecommand \bibfnamefont [1]{#1}%
\providecommand \citenamefont [1]{#1}%
\providecommand \href@noop [0]{\@secondoftwo}%
\providecommand \href [0]{\begingroup \@sanitize@url \@href}%
\providecommand \@href[1]{\@@startlink{#1}\@@href}%
\providecommand \@@href[1]{\endgroup#1\@@endlink}%
\providecommand \@sanitize@url [0]{\catcode `\\12\catcode `\$12\catcode
  `\&12\catcode `\#12\catcode `\^12\catcode `\_12\catcode `\%12\relax}%
\providecommand \@@startlink[1]{}%
\providecommand \@@endlink[0]{}%
\providecommand \url  [0]{\begingroup\@sanitize@url \@url }%
\providecommand \@url [1]{\endgroup\@href {#1}{\urlprefix }}%
\providecommand \urlprefix  [0]{URL }%
\providecommand \Eprint [0]{\href }%
\providecommand \doibase [0]{http://dx.doi.org/}%
\providecommand \selectlanguage [0]{\@gobble}%
\providecommand \bibinfo  [0]{\@secondoftwo}%
\providecommand \bibfield  [0]{\@secondoftwo}%
\providecommand \translation [1]{[#1]}%
\providecommand \BibitemOpen [0]{}%
\providecommand \bibitemStop [0]{}%
\providecommand \bibitemNoStop [0]{.\EOS\space}%
\providecommand \EOS [0]{\spacefactor3000\relax}%
\providecommand \BibitemShut  [1]{\csname bibitem#1\endcsname}%
\let\auto@bib@innerbib\@empty
\bibitem [{\citenamefont {Bistritzer}\ and\ \citenamefont
  {MacDonald}(2011)}]{bistritzer2011}%
  \BibitemOpen
  \bibfield  {author} {\bibinfo {author} {\bibfnamefont {R.}~\bibnamefont
  {Bistritzer}}\ and\ \bibinfo {author} {\bibfnamefont {A.~H.}\ \bibnamefont
  {MacDonald}},\ }\href@noop {} {\bibfield  {journal} {\bibinfo  {journal}
  {Proc. Natl. Acad. Sci.}\ }\textbf {\bibinfo {volume} {108}},\ \bibinfo
  {pages} {12233} (\bibinfo {year} {2011})}\BibitemShut {NoStop}%
\bibitem [{\citenamefont {Dos~Santos}\ \emph {et~al.}(2007)\citenamefont
  {Dos~Santos}, \citenamefont {Peres},\ and\ \citenamefont {Neto}}]{dos2007}%
  \BibitemOpen
  \bibfield  {author} {\bibinfo {author} {\bibfnamefont {J.~L.}\ \bibnamefont
  {Dos~Santos}}, \bibinfo {author} {\bibfnamefont {N.}~\bibnamefont {Peres}}, \
  and\ \bibinfo {author} {\bibfnamefont {A.~C.}\ \bibnamefont {Neto}},\
  }\href@noop {} {\bibfield  {journal} {\bibinfo  {journal} {Phys. Rev. Lett.}\
  }\textbf {\bibinfo {volume} {99}},\ \bibinfo {pages} {256802} (\bibinfo
  {year} {2007})}\BibitemShut {NoStop}%
\bibitem [{\citenamefont {Andrei}\ and\ \citenamefont
  {MacDonald}(2020)}]{andrei2020}%
  \BibitemOpen
  \bibfield  {author} {\bibinfo {author} {\bibfnamefont {E.~Y.}\ \bibnamefont
  {Andrei}}\ and\ \bibinfo {author} {\bibfnamefont {A.~H.}\ \bibnamefont
  {MacDonald}},\ }\href@noop {} {\bibfield  {journal} {\bibinfo  {journal}
  {Nat. Mat.}\ }\textbf {\bibinfo {volume} {19}},\ \bibinfo {pages} {1265}
  (\bibinfo {year} {2020})}\BibitemShut {NoStop}%
\bibitem [{\citenamefont {Balents}\ \emph {et~al.}(2020)\citenamefont
  {Balents}, \citenamefont {Dean}, \citenamefont {Efetov},\ and\ \citenamefont
  {Young}}]{balents2020}%
  \BibitemOpen
  \bibfield  {author} {\bibinfo {author} {\bibfnamefont {L.}~\bibnamefont
  {Balents}}, \bibinfo {author} {\bibfnamefont {C.~R.}\ \bibnamefont {Dean}},
  \bibinfo {author} {\bibfnamefont {D.~K.}\ \bibnamefont {Efetov}}, \ and\
  \bibinfo {author} {\bibfnamefont {A.~F.}\ \bibnamefont {Young}},\ }\href@noop
  {} {\bibfield  {journal} {\bibinfo  {journal} {Nat. Phys.}\ }\textbf
  {\bibinfo {volume} {16}},\ \bibinfo {pages} {725} (\bibinfo {year}
  {2020})}\BibitemShut {NoStop}%
\bibitem [{\citenamefont {T{\"o}rm{\"a}}\ \emph {et~al.}(2022)\citenamefont
  {T{\"o}rm{\"a}}, \citenamefont {Peotta},\ and\ \citenamefont
  {Bernevig}}]{torma2022}%
  \BibitemOpen
  \bibfield  {author} {\bibinfo {author} {\bibfnamefont {P.}~\bibnamefont
  {T{\"o}rm{\"a}}}, \bibinfo {author} {\bibfnamefont {S.}~\bibnamefont
  {Peotta}}, \ and\ \bibinfo {author} {\bibfnamefont {B.~A.}\ \bibnamefont
  {Bernevig}},\ }\href@noop {} {\bibfield  {journal} {\bibinfo  {journal} {Nat.
  Rev. Phys.}\ }\textbf {\bibinfo {volume} {4}},\ \bibinfo {pages} {528}
  (\bibinfo {year} {2022})}\BibitemShut {NoStop}%
\bibitem [{\citenamefont {Cao}\ \emph {et~al.}(2018{\natexlab{a}})\citenamefont
  {Cao}, \citenamefont {Fatemi}, \citenamefont {Fang}, \citenamefont
  {Watanabe}, \citenamefont {Taniguchi}, \citenamefont {Kaxiras},\ and\
  \citenamefont {Jarillo-Herrero}}]{cao2018a}%
  \BibitemOpen
  \bibfield  {author} {\bibinfo {author} {\bibfnamefont {Y.}~\bibnamefont
  {Cao}}, \bibinfo {author} {\bibfnamefont {V.}~\bibnamefont {Fatemi}},
  \bibinfo {author} {\bibfnamefont {S.}~\bibnamefont {Fang}}, \bibinfo {author}
  {\bibfnamefont {K.}~\bibnamefont {Watanabe}}, \bibinfo {author}
  {\bibfnamefont {T.}~\bibnamefont {Taniguchi}}, \bibinfo {author}
  {\bibfnamefont {E.}~\bibnamefont {Kaxiras}}, \ and\ \bibinfo {author}
  {\bibfnamefont {P.}~\bibnamefont {Jarillo-Herrero}},\ }\href@noop {}
  {\bibfield  {journal} {\bibinfo  {journal} {Nature}\ }\textbf {\bibinfo
  {volume} {556}},\ \bibinfo {pages} {43} (\bibinfo {year}
  {2018}{\natexlab{a}})}\BibitemShut {NoStop}%
\bibitem [{\citenamefont {Yankowitz}\ \emph
  {et~al.}(2019{\natexlab{a}})\citenamefont {Yankowitz}, \citenamefont {Chen},
  \citenamefont {Polshyn}, \citenamefont {Zhang}, \citenamefont {Watanabe},
  \citenamefont {Taniguchi}, \citenamefont {Graf}, \citenamefont {Young},\ and\
  \citenamefont {Dean}}]{yankowitz2019}%
  \BibitemOpen
  \bibfield  {author} {\bibinfo {author} {\bibfnamefont {M.}~\bibnamefont
  {Yankowitz}}, \bibinfo {author} {\bibfnamefont {S.}~\bibnamefont {Chen}},
  \bibinfo {author} {\bibfnamefont {H.}~\bibnamefont {Polshyn}}, \bibinfo
  {author} {\bibfnamefont {Y.}~\bibnamefont {Zhang}}, \bibinfo {author}
  {\bibfnamefont {K.}~\bibnamefont {Watanabe}}, \bibinfo {author}
  {\bibfnamefont {T.}~\bibnamefont {Taniguchi}}, \bibinfo {author}
  {\bibfnamefont {D.}~\bibnamefont {Graf}}, \bibinfo {author} {\bibfnamefont
  {A.~F.}\ \bibnamefont {Young}}, \ and\ \bibinfo {author} {\bibfnamefont
  {C.~R.}\ \bibnamefont {Dean}},\ }\href@noop {} {\bibfield  {journal}
  {\bibinfo  {journal} {Science}\ }\textbf {\bibinfo {volume} {363}},\ \bibinfo
  {pages} {1059} (\bibinfo {year} {2019}{\natexlab{a}})}\BibitemShut {NoStop}%
\bibitem [{\citenamefont {Oh}\ \emph {et~al.}(2021)\citenamefont {Oh},
  \citenamefont {Nuckolls}, \citenamefont {Wong}, \citenamefont {Lee},
  \citenamefont {Liu}, \citenamefont {Watanabe}, \citenamefont {Taniguchi},\
  and\ \citenamefont {Yazdani}}]{oh2021}%
  \BibitemOpen
  \bibfield  {author} {\bibinfo {author} {\bibfnamefont {M.}~\bibnamefont
  {Oh}}, \bibinfo {author} {\bibfnamefont {K.~P.}\ \bibnamefont {Nuckolls}},
  \bibinfo {author} {\bibfnamefont {D.}~\bibnamefont {Wong}}, \bibinfo {author}
  {\bibfnamefont {R.~L.}\ \bibnamefont {Lee}}, \bibinfo {author} {\bibfnamefont
  {X.}~\bibnamefont {Liu}}, \bibinfo {author} {\bibfnamefont {K.}~\bibnamefont
  {Watanabe}}, \bibinfo {author} {\bibfnamefont {T.}~\bibnamefont {Taniguchi}},
  \ and\ \bibinfo {author} {\bibfnamefont {A.}~\bibnamefont {Yazdani}},\
  }\href@noop {} {\bibfield  {journal} {\bibinfo  {journal} {Nature}\ }\textbf
  {\bibinfo {volume} {600}},\ \bibinfo {pages} {240} (\bibinfo {year}
  {2021})}\BibitemShut {NoStop}%
\bibitem [{\citenamefont {Yankowitz}\ \emph
  {et~al.}(2019{\natexlab{b}})\citenamefont {Yankowitz}, \citenamefont {Chen},
  \citenamefont {Polshyn}, \citenamefont {Zhang}, \citenamefont {Watanabe},
  \citenamefont {Taniguchi}, \citenamefont {Graf}, \citenamefont {Young},\ and\
  \citenamefont {Dean}}]{yankowitz2019tuning}%
  \BibitemOpen
  \bibfield  {author} {\bibinfo {author} {\bibfnamefont {M.}~\bibnamefont
  {Yankowitz}}, \bibinfo {author} {\bibfnamefont {S.}~\bibnamefont {Chen}},
  \bibinfo {author} {\bibfnamefont {H.}~\bibnamefont {Polshyn}}, \bibinfo
  {author} {\bibfnamefont {Y.}~\bibnamefont {Zhang}}, \bibinfo {author}
  {\bibfnamefont {K.}~\bibnamefont {Watanabe}}, \bibinfo {author}
  {\bibfnamefont {T.}~\bibnamefont {Taniguchi}}, \bibinfo {author}
  {\bibfnamefont {D.}~\bibnamefont {Graf}}, \bibinfo {author} {\bibfnamefont
  {A.~F.}\ \bibnamefont {Young}}, \ and\ \bibinfo {author} {\bibfnamefont
  {C.~R.}\ \bibnamefont {Dean}},\ }\href@noop {} {\bibfield  {journal}
  {\bibinfo  {journal} {Science}\ }\textbf {\bibinfo {volume} {363}},\ \bibinfo
  {pages} {1059} (\bibinfo {year} {2019}{\natexlab{b}})}\BibitemShut {NoStop}%
\bibitem [{\citenamefont {Cao}\ \emph {et~al.}(2018{\natexlab{b}})\citenamefont
  {Cao}, \citenamefont {Fatemi}, \citenamefont {Demir}, \citenamefont {Fang},
  \citenamefont {Tomarken}, \citenamefont {Luo}, \citenamefont
  {Sanchez-Yamagishi}, \citenamefont {Watanabe}, \citenamefont {Taniguchi},
  \citenamefont {Kaxiras} \emph {et~al.}}]{cao2018b}%
  \BibitemOpen
  \bibfield  {author} {\bibinfo {author} {\bibfnamefont {Y.}~\bibnamefont
  {Cao}}, \bibinfo {author} {\bibfnamefont {V.}~\bibnamefont {Fatemi}},
  \bibinfo {author} {\bibfnamefont {A.}~\bibnamefont {Demir}}, \bibinfo
  {author} {\bibfnamefont {S.}~\bibnamefont {Fang}}, \bibinfo {author}
  {\bibfnamefont {S.~L.}\ \bibnamefont {Tomarken}}, \bibinfo {author}
  {\bibfnamefont {J.~Y.}\ \bibnamefont {Luo}}, \bibinfo {author} {\bibfnamefont
  {J.~D.}\ \bibnamefont {Sanchez-Yamagishi}}, \bibinfo {author} {\bibfnamefont
  {K.}~\bibnamefont {Watanabe}}, \bibinfo {author} {\bibfnamefont
  {T.}~\bibnamefont {Taniguchi}}, \bibinfo {author} {\bibfnamefont
  {E.}~\bibnamefont {Kaxiras}},  \emph {et~al.},\ }\href@noop {} {\bibfield
  {journal} {\bibinfo  {journal} {Nature}\ }\textbf {\bibinfo {volume} {556}},\
  \bibinfo {pages} {80} (\bibinfo {year} {2018}{\natexlab{b}})}\BibitemShut
  {NoStop}%
\bibitem [{\citenamefont {Codecido}\ \emph {et~al.}(2019)\citenamefont
  {Codecido}, \citenamefont {Wang}, \citenamefont {Koester}, \citenamefont
  {Che}, \citenamefont {Tian}, \citenamefont {Lv}, \citenamefont {Tran},
  \citenamefont {Watanabe}, \citenamefont {Taniguchi}, \citenamefont {Zhang}
  \emph {et~al.}}]{codecido2019}%
  \BibitemOpen
  \bibfield  {author} {\bibinfo {author} {\bibfnamefont {E.}~\bibnamefont
  {Codecido}}, \bibinfo {author} {\bibfnamefont {Q.}~\bibnamefont {Wang}},
  \bibinfo {author} {\bibfnamefont {R.}~\bibnamefont {Koester}}, \bibinfo
  {author} {\bibfnamefont {S.}~\bibnamefont {Che}}, \bibinfo {author}
  {\bibfnamefont {H.}~\bibnamefont {Tian}}, \bibinfo {author} {\bibfnamefont
  {R.}~\bibnamefont {Lv}}, \bibinfo {author} {\bibfnamefont {S.}~\bibnamefont
  {Tran}}, \bibinfo {author} {\bibfnamefont {K.}~\bibnamefont {Watanabe}},
  \bibinfo {author} {\bibfnamefont {T.}~\bibnamefont {Taniguchi}}, \bibinfo
  {author} {\bibfnamefont {F.}~\bibnamefont {Zhang}},  \emph {et~al.},\
  }\href@noop {} {\bibfield  {journal} {\bibinfo  {journal} {Science Adv.}\
  }\textbf {\bibinfo {volume} {5}},\ \bibinfo {pages} {eaaw9770} (\bibinfo
  {year} {2019})}\BibitemShut {NoStop}%
\bibitem [{\citenamefont {Nuckolls}\ \emph {et~al.}(2020)\citenamefont
  {Nuckolls}, \citenamefont {Oh}, \citenamefont {Wong}, \citenamefont {Lian},
  \citenamefont {Watanabe}, \citenamefont {Taniguchi}, \citenamefont
  {Bernevig},\ and\ \citenamefont {Yazdani}}]{nuckolls2020}%
  \BibitemOpen
  \bibfield  {author} {\bibinfo {author} {\bibfnamefont {K.~P.}\ \bibnamefont
  {Nuckolls}}, \bibinfo {author} {\bibfnamefont {M.}~\bibnamefont {Oh}},
  \bibinfo {author} {\bibfnamefont {D.}~\bibnamefont {Wong}}, \bibinfo {author}
  {\bibfnamefont {B.}~\bibnamefont {Lian}}, \bibinfo {author} {\bibfnamefont
  {K.}~\bibnamefont {Watanabe}}, \bibinfo {author} {\bibfnamefont
  {T.}~\bibnamefont {Taniguchi}}, \bibinfo {author} {\bibfnamefont {B.~A.}\
  \bibnamefont {Bernevig}}, \ and\ \bibinfo {author} {\bibfnamefont
  {A.}~\bibnamefont {Yazdani}},\ }\href@noop {} {\bibfield  {journal} {\bibinfo
   {journal} {Nature}\ }\textbf {\bibinfo {volume} {588}},\ \bibinfo {pages}
  {610} (\bibinfo {year} {2020})}\BibitemShut {NoStop}%
\bibitem [{\citenamefont {Cao}\ \emph {et~al.}(2020)\citenamefont {Cao},
  \citenamefont {Rodan-Legrain}, \citenamefont {Rubies-Bigorda}, \citenamefont
  {Park}, \citenamefont {Watanabe}, \citenamefont {Taniguchi},\ and\
  \citenamefont {Jarillo-Herrero}}]{cao2020tunable}%
  \BibitemOpen
  \bibfield  {author} {\bibinfo {author} {\bibfnamefont {Y.}~\bibnamefont
  {Cao}}, \bibinfo {author} {\bibfnamefont {D.}~\bibnamefont {Rodan-Legrain}},
  \bibinfo {author} {\bibfnamefont {O.}~\bibnamefont {Rubies-Bigorda}},
  \bibinfo {author} {\bibfnamefont {J.~M.}\ \bibnamefont {Park}}, \bibinfo
  {author} {\bibfnamefont {K.}~\bibnamefont {Watanabe}}, \bibinfo {author}
  {\bibfnamefont {T.}~\bibnamefont {Taniguchi}}, \ and\ \bibinfo {author}
  {\bibfnamefont {P.}~\bibnamefont {Jarillo-Herrero}},\ }\href@noop {}
  {\bibfield  {journal} {\bibinfo  {journal} {Nature}\ }\textbf {\bibinfo
  {volume} {583}},\ \bibinfo {pages} {215} (\bibinfo {year}
  {2020})}\BibitemShut {NoStop}%
\bibitem [{\citenamefont {Trambly~de Laissardi{\`e}re}\ \emph
  {et~al.}(2010{\natexlab{a}})\citenamefont {Trambly~de Laissardi{\`e}re},
  \citenamefont {Mayou},\ and\ \citenamefont {Magaud}}]{trambly2010}%
  \BibitemOpen
  \bibfield  {author} {\bibinfo {author} {\bibfnamefont {G.}~\bibnamefont
  {Trambly~de Laissardi{\`e}re}}, \bibinfo {author} {\bibfnamefont
  {D.}~\bibnamefont {Mayou}}, \ and\ \bibinfo {author} {\bibfnamefont
  {L.}~\bibnamefont {Magaud}},\ }\href@noop {} {\bibfield  {journal} {\bibinfo
  {journal} {Nano Lett.}\ }\textbf {\bibinfo {volume} {10}},\ \bibinfo {pages}
  {804} (\bibinfo {year} {2010}{\natexlab{a}})}\BibitemShut {NoStop}%
\bibitem [{\citenamefont {Lisi}\ \emph
  {et~al.}(2021{\natexlab{a}})\citenamefont {Lisi}, \citenamefont {Lu},
  \citenamefont {Benschop}, \citenamefont {de~Jong}, \citenamefont {Stepanov},
  \citenamefont {Duran}, \citenamefont {Margot}, \citenamefont {Cucchi},
  \citenamefont {Cappelli}, \citenamefont {Hunter} \emph
  {et~al.}}]{lisi2021observation}%
  \BibitemOpen
  \bibfield  {author} {\bibinfo {author} {\bibfnamefont {S.}~\bibnamefont
  {Lisi}}, \bibinfo {author} {\bibfnamefont {X.}~\bibnamefont {Lu}}, \bibinfo
  {author} {\bibfnamefont {T.}~\bibnamefont {Benschop}}, \bibinfo {author}
  {\bibfnamefont {T.~A.}\ \bibnamefont {de~Jong}}, \bibinfo {author}
  {\bibfnamefont {P.}~\bibnamefont {Stepanov}}, \bibinfo {author}
  {\bibfnamefont {J.~R.}\ \bibnamefont {Duran}}, \bibinfo {author}
  {\bibfnamefont {F.}~\bibnamefont {Margot}}, \bibinfo {author} {\bibfnamefont
  {I.}~\bibnamefont {Cucchi}}, \bibinfo {author} {\bibfnamefont
  {E.}~\bibnamefont {Cappelli}}, \bibinfo {author} {\bibfnamefont
  {A.}~\bibnamefont {Hunter}},  \emph {et~al.},\ }\href@noop {} {\bibfield
  {journal} {\bibinfo  {journal} {Nat. Phys.}\ }\textbf {\bibinfo {volume}
  {17}},\ \bibinfo {pages} {189} (\bibinfo {year}
  {2021}{\natexlab{a}})}\BibitemShut {NoStop}%
\bibitem [{\citenamefont {Morell}\ \emph {et~al.}(2010)\citenamefont {Morell},
  \citenamefont {Correa}, \citenamefont {Vargas}, \citenamefont {Pacheco},\
  and\ \citenamefont {Barticevic}}]{morell2010}%
  \BibitemOpen
  \bibfield  {author} {\bibinfo {author} {\bibfnamefont {E.~S.}\ \bibnamefont
  {Morell}}, \bibinfo {author} {\bibfnamefont {J.}~\bibnamefont {Correa}},
  \bibinfo {author} {\bibfnamefont {P.}~\bibnamefont {Vargas}}, \bibinfo
  {author} {\bibfnamefont {M.}~\bibnamefont {Pacheco}}, \ and\ \bibinfo
  {author} {\bibfnamefont {Z.}~\bibnamefont {Barticevic}},\ }\href@noop {}
  {\bibfield  {journal} {\bibinfo  {journal} {Phys. Rev. B}\ }\textbf {\bibinfo
  {volume} {82}},\ \bibinfo {pages} {121407} (\bibinfo {year}
  {2010})}\BibitemShut {NoStop}%
\bibitem [{\citenamefont {Tarnopolsky}\ \emph {et~al.}(2019)\citenamefont
  {Tarnopolsky}, \citenamefont {Kruchkov},\ and\ \citenamefont
  {Vishwanath}}]{tarnopolsky2019}%
  \BibitemOpen
  \bibfield  {author} {\bibinfo {author} {\bibfnamefont {G.}~\bibnamefont
  {Tarnopolsky}}, \bibinfo {author} {\bibfnamefont {A.~J.}\ \bibnamefont
  {Kruchkov}}, \ and\ \bibinfo {author} {\bibfnamefont {A.}~\bibnamefont
  {Vishwanath}},\ }\href@noop {} {\bibfield  {journal} {\bibinfo  {journal}
  {Phys. Rev. Lett.}\ }\textbf {\bibinfo {volume} {122}},\ \bibinfo {pages}
  {106405} (\bibinfo {year} {2019})}\BibitemShut {NoStop}%
\bibitem [{\citenamefont {Lisi}\ \emph
  {et~al.}(2021{\natexlab{b}})\citenamefont {Lisi}, \citenamefont {Lu},
  \citenamefont {Benschop}, \citenamefont {de~Jong}, \citenamefont {Stepanov},
  \citenamefont {Duran}, \citenamefont {Margot}, \citenamefont {Cucchi},
  \citenamefont {Cappelli}, \citenamefont {Hunter} \emph {et~al.}}]{lisi2021}%
  \BibitemOpen
  \bibfield  {author} {\bibinfo {author} {\bibfnamefont {S.}~\bibnamefont
  {Lisi}}, \bibinfo {author} {\bibfnamefont {X.}~\bibnamefont {Lu}}, \bibinfo
  {author} {\bibfnamefont {T.}~\bibnamefont {Benschop}}, \bibinfo {author}
  {\bibfnamefont {T.~A.}\ \bibnamefont {de~Jong}}, \bibinfo {author}
  {\bibfnamefont {P.}~\bibnamefont {Stepanov}}, \bibinfo {author}
  {\bibfnamefont {J.~R.}\ \bibnamefont {Duran}}, \bibinfo {author}
  {\bibfnamefont {F.}~\bibnamefont {Margot}}, \bibinfo {author} {\bibfnamefont
  {I.}~\bibnamefont {Cucchi}}, \bibinfo {author} {\bibfnamefont
  {E.}~\bibnamefont {Cappelli}}, \bibinfo {author} {\bibfnamefont
  {A.}~\bibnamefont {Hunter}},  \emph {et~al.},\ }\href@noop {} {\bibfield
  {journal} {\bibinfo  {journal} {Nat. Phys.}\ }\textbf {\bibinfo {volume}
  {17}},\ \bibinfo {pages} {189} (\bibinfo {year}
  {2021}{\natexlab{b}})}\BibitemShut {NoStop}%
\bibitem [{\citenamefont {Gonz{\'a}lez-Tudela}\ and\ \citenamefont
  {Cirac}(2019)}]{gonzalez2019}%
  \BibitemOpen
  \bibfield  {author} {\bibinfo {author} {\bibfnamefont {A.}~\bibnamefont
  {Gonz{\'a}lez-Tudela}}\ and\ \bibinfo {author} {\bibfnamefont {J.~I.}\
  \bibnamefont {Cirac}},\ }\href@noop {} {\bibfield  {journal} {\bibinfo
  {journal} {Phys. Rev. A}\ }\textbf {\bibinfo {volume} {100}},\ \bibinfo
  {pages} {053604} (\bibinfo {year} {2019})}\BibitemShut {NoStop}%
\bibitem [{\citenamefont {Salamon}\ \emph {et~al.}(2020)\citenamefont
  {Salamon}, \citenamefont {Celi}, \citenamefont {Chhajlany}, \citenamefont
  {Fr{\'e}rot}, \citenamefont {Lewenstein}, \citenamefont {Tarruell},\ and\
  \citenamefont {Rakshit}}]{salamon2020}%
  \BibitemOpen
  \bibfield  {author} {\bibinfo {author} {\bibfnamefont {T.}~\bibnamefont
  {Salamon}}, \bibinfo {author} {\bibfnamefont {A.}~\bibnamefont {Celi}},
  \bibinfo {author} {\bibfnamefont {R.~W.}\ \bibnamefont {Chhajlany}}, \bibinfo
  {author} {\bibfnamefont {I.}~\bibnamefont {Fr{\'e}rot}}, \bibinfo {author}
  {\bibfnamefont {M.}~\bibnamefont {Lewenstein}}, \bibinfo {author}
  {\bibfnamefont {L.}~\bibnamefont {Tarruell}}, \ and\ \bibinfo {author}
  {\bibfnamefont {D.}~\bibnamefont {Rakshit}},\ }\href@noop {} {\bibfield
  {journal} {\bibinfo  {journal} {Phys. Rev. Lett.}\ }\textbf {\bibinfo
  {volume} {125}},\ \bibinfo {pages} {030504} (\bibinfo {year}
  {2020})}\BibitemShut {NoStop}%
\bibitem [{\citenamefont {Luo}\ and\ \citenamefont
  {Zhang}(2021)}]{luo2021spin}%
  \BibitemOpen
  \bibfield  {author} {\bibinfo {author} {\bibfnamefont {X.-W.}\ \bibnamefont
  {Luo}}\ and\ \bibinfo {author} {\bibfnamefont {C.}~\bibnamefont {Zhang}},\
  }\href@noop {} {\bibfield  {journal} {\bibinfo  {journal} {Phys. Rev. Lett.}\
  }\textbf {\bibinfo {volume} {126}},\ \bibinfo {pages} {103201} (\bibinfo
  {year} {2021})}\BibitemShut {NoStop}%
\bibitem [{\citenamefont {Lee}\ and\ \citenamefont {Pixley}(2022)}]{lee2022}%
  \BibitemOpen
  \bibfield  {author} {\bibinfo {author} {\bibfnamefont {J.}~\bibnamefont
  {Lee}}\ and\ \bibinfo {author} {\bibfnamefont {J.}~\bibnamefont {Pixley}},\
  }\href@noop {} {\bibfield  {journal} {\bibinfo  {journal} {SciPost Phys.}\
  }\textbf {\bibinfo {volume} {13}},\ \bibinfo {pages} {033} (\bibinfo {year}
  {2022})}\BibitemShut {NoStop}%
\bibitem [{\citenamefont {Meng}\ \emph {et~al.}(2023)\citenamefont {Meng},
  \citenamefont {Wang}, \citenamefont {Han}, \citenamefont {Liu}, \citenamefont
  {Wen}, \citenamefont {Gao}, \citenamefont {Wang}, \citenamefont {Chin},\ and\
  \citenamefont {Zhang}}]{meng2023atomic}%
  \BibitemOpen
  \bibfield  {author} {\bibinfo {author} {\bibfnamefont {Z.}~\bibnamefont
  {Meng}}, \bibinfo {author} {\bibfnamefont {L.}~\bibnamefont {Wang}}, \bibinfo
  {author} {\bibfnamefont {W.}~\bibnamefont {Han}}, \bibinfo {author}
  {\bibfnamefont {F.}~\bibnamefont {Liu}}, \bibinfo {author} {\bibfnamefont
  {K.}~\bibnamefont {Wen}}, \bibinfo {author} {\bibfnamefont {C.}~\bibnamefont
  {Gao}}, \bibinfo {author} {\bibfnamefont {P.}~\bibnamefont {Wang}}, \bibinfo
  {author} {\bibfnamefont {C.}~\bibnamefont {Chin}}, \ and\ \bibinfo {author}
  {\bibfnamefont {J.}~\bibnamefont {Zhang}},\ }\href@noop {} {\bibfield
  {journal} {\bibinfo  {journal} {Nature}\ }\textbf {\bibinfo {volume} {615}},\
  \bibinfo {pages} {231} (\bibinfo {year} {2023})}\BibitemShut {NoStop}%
\bibitem [{\citenamefont {Aubry}\ and\ \citenamefont
  {Andr{\'e}}(1980)}]{aubry1980}%
  \BibitemOpen
  \bibfield  {author} {\bibinfo {author} {\bibfnamefont {S.}~\bibnamefont
  {Aubry}}\ and\ \bibinfo {author} {\bibfnamefont {G.}~\bibnamefont
  {Andr{\'e}}},\ }\href@noop {} {\bibfield  {journal} {\bibinfo  {journal}
  {Ann. Israel Phys. Soc}\ }\textbf {\bibinfo {volume} {3}},\ \bibinfo {pages}
  {18} (\bibinfo {year} {1980})}\BibitemShut {NoStop}%
\bibitem [{\citenamefont {Roati}\ \emph {et~al.}(2008)\citenamefont {Roati},
  \citenamefont {D'Errico}, \citenamefont {Fallani}, \citenamefont {Fattori},
  \citenamefont {Fort}, \citenamefont {Zaccanti}, \citenamefont {Modugno},
  \citenamefont {Modugno},\ and\ \citenamefont {Inguscio}}]{Roati2008}%
  \BibitemOpen
  \bibfield  {author} {\bibinfo {author} {\bibfnamefont {G.}~\bibnamefont
  {Roati}}, \bibinfo {author} {\bibfnamefont {C.}~\bibnamefont {D'Errico}},
  \bibinfo {author} {\bibfnamefont {L.}~\bibnamefont {Fallani}}, \bibinfo
  {author} {\bibfnamefont {M.}~\bibnamefont {Fattori}}, \bibinfo {author}
  {\bibfnamefont {C.}~\bibnamefont {Fort}}, \bibinfo {author} {\bibfnamefont
  {M.}~\bibnamefont {Zaccanti}}, \bibinfo {author} {\bibfnamefont
  {G.}~\bibnamefont {Modugno}}, \bibinfo {author} {\bibfnamefont
  {M.}~\bibnamefont {Modugno}}, \ and\ \bibinfo {author} {\bibfnamefont
  {M.}~\bibnamefont {Inguscio}},\ }\href@noop {} {\bibfield  {journal}
  {\bibinfo  {journal} {Nature}\ }\textbf {\bibinfo {volume} {453}},\ \bibinfo
  {pages} {895} (\bibinfo {year} {2008})}\BibitemShut {NoStop}%
\bibitem [{\citenamefont {Schreiber}\ \emph {et~al.}(2015)\citenamefont
  {Schreiber}, \citenamefont {Hodgman}, \citenamefont {Bordia}, \citenamefont
  {Lüschen}, \citenamefont {Fischer}, \citenamefont {Vosk}, \citenamefont
  {Altman}, \citenamefont {Schneider},\ and\ \citenamefont
  {Bloch}}]{Schreiber2015}%
  \BibitemOpen
  \bibfield  {author} {\bibinfo {author} {\bibfnamefont {M.}~\bibnamefont
  {Schreiber}}, \bibinfo {author} {\bibfnamefont {S.~S.}\ \bibnamefont
  {Hodgman}}, \bibinfo {author} {\bibfnamefont {P.}~\bibnamefont {Bordia}},
  \bibinfo {author} {\bibfnamefont {H.~P.}\ \bibnamefont {Lüschen}}, \bibinfo
  {author} {\bibfnamefont {M.~H.}\ \bibnamefont {Fischer}}, \bibinfo {author}
  {\bibfnamefont {R.}~\bibnamefont {Vosk}}, \bibinfo {author} {\bibfnamefont
  {E.}~\bibnamefont {Altman}}, \bibinfo {author} {\bibfnamefont
  {U.}~\bibnamefont {Schneider}}, \ and\ \bibinfo {author} {\bibfnamefont
  {I.}~\bibnamefont {Bloch}},\ }\href@noop {} {\bibfield  {journal} {\bibinfo
  {journal} {Science}\ }\textbf {\bibinfo {volume} {349}},\ \bibinfo {pages}
  {842} (\bibinfo {year} {2015})}\BibitemShut {NoStop}%
\bibitem [{\citenamefont {Viebahn}\ \emph {et~al.}(2019)\citenamefont
  {Viebahn}, \citenamefont {Sbroscia}, \citenamefont {Carter}, \citenamefont
  {Yu},\ and\ \citenamefont {Schneider}}]{Viebahn2019}%
  \BibitemOpen
  \bibfield  {author} {\bibinfo {author} {\bibfnamefont {K.}~\bibnamefont
  {Viebahn}}, \bibinfo {author} {\bibfnamefont {M.}~\bibnamefont {Sbroscia}},
  \bibinfo {author} {\bibfnamefont {E.}~\bibnamefont {Carter}}, \bibinfo
  {author} {\bibfnamefont {J.-C.}\ \bibnamefont {Yu}}, \ and\ \bibinfo {author}
  {\bibfnamefont {U.}~\bibnamefont {Schneider}},\ }\href@noop {} {\bibfield
  {journal} {\bibinfo  {journal} {Phys. Rev. Lett.}\ }\textbf {\bibinfo
  {volume} {122}},\ \bibinfo {pages} {110404} (\bibinfo {year}
  {2019})}\BibitemShut {NoStop}%
\bibitem [{\citenamefont {Lahini}\ \emph {et~al.}(2009)\citenamefont {Lahini},
  \citenamefont {Pugatch}, \citenamefont {Pozzi}, \citenamefont {Sorel},
  \citenamefont {Morandotti}, \citenamefont {Davidson},\ and\ \citenamefont
  {Silberberg}}]{lahini2009}%
  \BibitemOpen
  \bibfield  {author} {\bibinfo {author} {\bibfnamefont {Y.}~\bibnamefont
  {Lahini}}, \bibinfo {author} {\bibfnamefont {R.}~\bibnamefont {Pugatch}},
  \bibinfo {author} {\bibfnamefont {F.}~\bibnamefont {Pozzi}}, \bibinfo
  {author} {\bibfnamefont {M.}~\bibnamefont {Sorel}}, \bibinfo {author}
  {\bibfnamefont {R.}~\bibnamefont {Morandotti}}, \bibinfo {author}
  {\bibfnamefont {N.}~\bibnamefont {Davidson}}, \ and\ \bibinfo {author}
  {\bibfnamefont {Y.}~\bibnamefont {Silberberg}},\ }\href@noop {} {\bibfield
  {journal} {\bibinfo  {journal} {Phys. Rev. lett.}\ }\textbf {\bibinfo
  {volume} {103}},\ \bibinfo {pages} {013901} (\bibinfo {year}
  {2009})}\BibitemShut {NoStop}%
\bibitem [{\citenamefont {Wang}\ \emph {et~al.}(2020)\citenamefont {Wang},
  \citenamefont {Zheng}, \citenamefont {Chen}, \citenamefont {Huang},
  \citenamefont {Kartashov}, \citenamefont {Torner}, \citenamefont {Konotop},\
  and\ \citenamefont {Ye}}]{wang2020}%
  \BibitemOpen
  \bibfield  {author} {\bibinfo {author} {\bibfnamefont {P.}~\bibnamefont
  {Wang}}, \bibinfo {author} {\bibfnamefont {Y.}~\bibnamefont {Zheng}},
  \bibinfo {author} {\bibfnamefont {X.}~\bibnamefont {Chen}}, \bibinfo {author}
  {\bibfnamefont {C.}~\bibnamefont {Huang}}, \bibinfo {author} {\bibfnamefont
  {Y.~V.}\ \bibnamefont {Kartashov}}, \bibinfo {author} {\bibfnamefont
  {L.}~\bibnamefont {Torner}}, \bibinfo {author} {\bibfnamefont {V.~V.}\
  \bibnamefont {Konotop}}, \ and\ \bibinfo {author} {\bibfnamefont
  {F.}~\bibnamefont {Ye}},\ }\href@noop {} {\bibfield  {journal} {\bibinfo
  {journal} {Nature}\ }\textbf {\bibinfo {volume} {577}},\ \bibinfo {pages}
  {42} (\bibinfo {year} {2020})}\BibitemShut {NoStop}%
\bibitem [{\citenamefont {Huang}\ \emph {et~al.}(2016)\citenamefont {Huang},
  \citenamefont {Ye}, \citenamefont {Chen}, \citenamefont {Kartashov},
  \citenamefont {Konotop},\ and\ \citenamefont {Torner}}]{huang2016}%
  \BibitemOpen
  \bibfield  {author} {\bibinfo {author} {\bibfnamefont {C.}~\bibnamefont
  {Huang}}, \bibinfo {author} {\bibfnamefont {F.}~\bibnamefont {Ye}}, \bibinfo
  {author} {\bibfnamefont {X.}~\bibnamefont {Chen}}, \bibinfo {author}
  {\bibfnamefont {Y.~V.}\ \bibnamefont {Kartashov}}, \bibinfo {author}
  {\bibfnamefont {V.~V.}\ \bibnamefont {Konotop}}, \ and\ \bibinfo {author}
  {\bibfnamefont {L.}~\bibnamefont {Torner}},\ }\href@noop {} {\bibfield
  {journal} {\bibinfo  {journal} {Scientific reports}\ }\textbf {\bibinfo
  {volume} {6}},\ \bibinfo {pages} {1} (\bibinfo {year} {2016})}\BibitemShut
  {NoStop}%
\bibitem [{\citenamefont {Trambly~de Laissardi{\`e}re}\ \emph
  {et~al.}(2010{\natexlab{b}})\citenamefont {Trambly~de Laissardi{\`e}re},
  \citenamefont {Mayou},\ and\ \citenamefont
  {Magaud}}]{trambly2010localization}%
  \BibitemOpen
  \bibfield  {author} {\bibinfo {author} {\bibfnamefont {G.}~\bibnamefont
  {Trambly~de Laissardi{\`e}re}}, \bibinfo {author} {\bibfnamefont
  {D.}~\bibnamefont {Mayou}}, \ and\ \bibinfo {author} {\bibfnamefont
  {L.}~\bibnamefont {Magaud}},\ }\href@noop {} {\bibfield  {journal} {\bibinfo
  {journal} {Nano lett.}\ }\textbf {\bibinfo {volume} {10}},\ \bibinfo {pages}
  {804} (\bibinfo {year} {2010}{\natexlab{b}})}\BibitemShut {NoStop}%
\bibitem [{\citenamefont {Wegner}(1980)}]{wegner1980}%
  \BibitemOpen
  \bibfield  {author} {\bibinfo {author} {\bibfnamefont {F.}~\bibnamefont
  {Wegner}},\ }\href@noop {} {\bibfield  {journal} {\bibinfo  {journal} {Z.
  Phys. B Cond. Matt.}\ }\textbf {\bibinfo {volume} {36}},\ \bibinfo {pages}
  {209} (\bibinfo {year} {1980})}\BibitemShut {NoStop}%
\bibitem [{\citenamefont {Evers}\ and\ \citenamefont
  {Mirlin}(2000)}]{evers2000}%
  \BibitemOpen
  \bibfield  {author} {\bibinfo {author} {\bibfnamefont {F.}~\bibnamefont
  {Evers}}\ and\ \bibinfo {author} {\bibfnamefont {A.}~\bibnamefont {Mirlin}},\
  }\href@noop {} {\bibfield  {journal} {\bibinfo  {journal} {Phys. Rev. Lett.}\
  }\textbf {\bibinfo {volume} {84}},\ \bibinfo {pages} {3690} (\bibinfo {year}
  {2000})}\BibitemShut {NoStop}%
\bibitem [{\citenamefont {Murphy}\ \emph {et~al.}(2011)\citenamefont {Murphy},
  \citenamefont {Wortis},\ and\ \citenamefont {Atkinson}}]{murphy2011}%
  \BibitemOpen
  \bibfield  {author} {\bibinfo {author} {\bibfnamefont {N.}~\bibnamefont
  {Murphy}}, \bibinfo {author} {\bibfnamefont {R.}~\bibnamefont {Wortis}}, \
  and\ \bibinfo {author} {\bibfnamefont {W.}~\bibnamefont {Atkinson}},\
  }\href@noop {} {\bibfield  {journal} {\bibinfo  {journal} {Phys. Rev. B}\
  }\textbf {\bibinfo {volume} {83}},\ \bibinfo {pages} {184206} (\bibinfo
  {year} {2011})}\BibitemShut {NoStop}%
\bibitem [{\citenamefont {Jagannathan}(2021)}]{jagannathan2021fibonacci}%
  \BibitemOpen
  \bibfield  {author} {\bibinfo {author} {\bibfnamefont {A.}~\bibnamefont
  {Jagannathan}},\ }\href@noop {} {\bibfield  {journal} {\bibinfo  {journal}
  {Rev. Mod. Phys.}\ }\textbf {\bibinfo {volume} {93}},\ \bibinfo {pages}
  {045001} (\bibinfo {year} {2021})}\BibitemShut {NoStop}%
\bibitem [{\citenamefont {Pook}\ and\ \citenamefont
  {Jan{\ss}en}(1991)}]{pook1991}%
  \BibitemOpen
  \bibfield  {author} {\bibinfo {author} {\bibfnamefont {W.}~\bibnamefont
  {Pook}}\ and\ \bibinfo {author} {\bibfnamefont {M.}~\bibnamefont
  {Jan{\ss}en}},\ }\href@noop {} {\bibfield  {journal} {\bibinfo  {journal} {Z.
  Phys. B Cond. Matt.}\ }\textbf {\bibinfo {volume} {82}},\ \bibinfo {pages}
  {295} (\bibinfo {year} {1991})}\BibitemShut {NoStop}%
\bibitem [{\citenamefont {Chamon}\ \emph {et~al.}(1996)\citenamefont {Chamon},
  \citenamefont {Mudry},\ and\ \citenamefont {Wen}}]{chamon1996}%
  \BibitemOpen
  \bibfield  {author} {\bibinfo {author} {\bibfnamefont {C.~d.~C.}\
  \bibnamefont {Chamon}}, \bibinfo {author} {\bibfnamefont {C.}~\bibnamefont
  {Mudry}}, \ and\ \bibinfo {author} {\bibfnamefont {X.-G.}\ \bibnamefont
  {Wen}},\ }\href@noop {} {\bibfield  {journal} {\bibinfo  {journal} {Phys.
  Rev. Lett.}\ }\textbf {\bibinfo {volume} {77}},\ \bibinfo {pages} {4194}
  (\bibinfo {year} {1996})}\BibitemShut {NoStop}%
\bibitem [{\citenamefont {Deng}\ \emph {et~al.}(2016)\citenamefont {Deng},
  \citenamefont {Altshuler}, \citenamefont {Shlyapnikov},\ and\ \citenamefont
  {Santos}}]{deng2016}%
  \BibitemOpen
  \bibfield  {author} {\bibinfo {author} {\bibfnamefont {X.}~\bibnamefont
  {Deng}}, \bibinfo {author} {\bibfnamefont {B.}~\bibnamefont {Altshuler}},
  \bibinfo {author} {\bibfnamefont {G.}~\bibnamefont {Shlyapnikov}}, \ and\
  \bibinfo {author} {\bibfnamefont {L.}~\bibnamefont {Santos}},\ }\href@noop {}
  {\bibfield  {journal} {\bibinfo  {journal} {Phys. Rev. Lett.}\ }\textbf
  {\bibinfo {volume} {117}},\ \bibinfo {pages} {020401} (\bibinfo {year}
  {2016})}\BibitemShut {NoStop}%
\bibitem [{\citenamefont {Mac{\'e}}\ \emph {et~al.}(2019)\citenamefont
  {Mac{\'e}}, \citenamefont {Alet},\ and\ \citenamefont
  {Laflorencie}}]{mace2019}%
  \BibitemOpen
  \bibfield  {author} {\bibinfo {author} {\bibfnamefont {N.}~\bibnamefont
  {Mac{\'e}}}, \bibinfo {author} {\bibfnamefont {F.}~\bibnamefont {Alet}}, \
  and\ \bibinfo {author} {\bibfnamefont {N.}~\bibnamefont {Laflorencie}},\
  }\href@noop {} {\bibfield  {journal} {\bibinfo  {journal} {Phys. Rev. Lett.}\
  }\textbf {\bibinfo {volume} {123}},\ \bibinfo {pages} {180601} (\bibinfo
  {year} {2019})}\BibitemShut {NoStop}%
\bibitem [{\citenamefont {Deng}\ \emph {et~al.}(2019)\citenamefont {Deng},
  \citenamefont {Ray}, \citenamefont {Sinha}, \citenamefont {Shlyapnikov},\
  and\ \citenamefont {Santos}}]{deng2019one}%
  \BibitemOpen
  \bibfield  {author} {\bibinfo {author} {\bibfnamefont {X.}~\bibnamefont
  {Deng}}, \bibinfo {author} {\bibfnamefont {S.}~\bibnamefont {Ray}}, \bibinfo
  {author} {\bibfnamefont {S.}~\bibnamefont {Sinha}}, \bibinfo {author}
  {\bibfnamefont {G.}~\bibnamefont {Shlyapnikov}}, \ and\ \bibinfo {author}
  {\bibfnamefont {L.}~\bibnamefont {Santos}},\ }\href@noop {} {\bibfield
  {journal} {\bibinfo  {journal} {Phys. Rev. Lett.}\ }\textbf {\bibinfo
  {volume} {123}},\ \bibinfo {pages} {025301} (\bibinfo {year}
  {2019})}\BibitemShut {NoStop}%
\bibitem [{\citenamefont {Sarkar}\ \emph {et~al.}(2022)\citenamefont {Sarkar},
  \citenamefont {Ghosh}, \citenamefont {Sen},\ and\ \citenamefont
  {Sengupta}}]{sarkar2022}%
  \BibitemOpen
  \bibfield  {author} {\bibinfo {author} {\bibfnamefont {M.}~\bibnamefont
  {Sarkar}}, \bibinfo {author} {\bibfnamefont {R.}~\bibnamefont {Ghosh}},
  \bibinfo {author} {\bibfnamefont {A.}~\bibnamefont {Sen}}, \ and\ \bibinfo
  {author} {\bibfnamefont {K.}~\bibnamefont {Sengupta}},\ }\href@noop {}
  {\bibfield  {journal} {\bibinfo  {journal} {Phys. Rev. B}\ }\textbf {\bibinfo
  {volume} {105}},\ \bibinfo {pages} {024301} (\bibinfo {year}
  {2022})}\BibitemShut {NoStop}%
\bibitem [{\citenamefont {Reisner}\ \emph {et~al.}(2022)\citenamefont
  {Reisner}, \citenamefont {Tahmi}, \citenamefont {Pi{\'e}chon}, \citenamefont
  {Kuhl},\ and\ \citenamefont {Mortessagne}}]{reisner2022experimental}%
  \BibitemOpen
  \bibfield  {author} {\bibinfo {author} {\bibfnamefont {M.}~\bibnamefont
  {Reisner}}, \bibinfo {author} {\bibfnamefont {Y.}~\bibnamefont {Tahmi}},
  \bibinfo {author} {\bibfnamefont {F.}~\bibnamefont {Pi{\'e}chon}}, \bibinfo
  {author} {\bibfnamefont {U.}~\bibnamefont {Kuhl}}, \ and\ \bibinfo {author}
  {\bibfnamefont {F.}~\bibnamefont {Mortessagne}},\ }\href@noop {} {\bibfield
  {journal} {\bibinfo  {journal} {arXiv preprint arXiv:2207.13755}\ } (\bibinfo
  {year} {2022})}\BibitemShut {NoStop}%
\bibitem [{\citenamefont {De~Luca}\ \emph {et~al.}(2014)\citenamefont
  {De~Luca}, \citenamefont {Altshuler}, \citenamefont {Kravtsov},\ and\
  \citenamefont {Scardicchio}}]{de2014anderson}%
  \BibitemOpen
  \bibfield  {author} {\bibinfo {author} {\bibfnamefont {A.}~\bibnamefont
  {De~Luca}}, \bibinfo {author} {\bibfnamefont {B.}~\bibnamefont {Altshuler}},
  \bibinfo {author} {\bibfnamefont {V.}~\bibnamefont {Kravtsov}}, \ and\
  \bibinfo {author} {\bibfnamefont {A.}~\bibnamefont {Scardicchio}},\
  }\href@noop {} {\bibfield  {journal} {\bibinfo  {journal} {Phys. Rev. Lett.}\
  }\textbf {\bibinfo {volume} {113}},\ \bibinfo {pages} {046806} (\bibinfo
  {year} {2014})}\BibitemShut {NoStop}%
\end{thebibliography}%

\end{document}